\documentclass[aps,pra,
,reprint,
superscriptaddress,
showpacs,preprintnumbers,
 amsmath,amssymb,
 aps,
]{revtex4-1}
\usepackage{xcolor}
\usepackage{graphicx}
\usepackage{dcolumn}
\usepackage{bm}
\usepackage{outlines}
\usepackage{float}
\usepackage{textcomp}
\usepackage{hyperref}
\let\vec=\mathbf

\usepackage[normalem]{ulem}

\begin{document}


\title{Bogoliubov-Cherenkov Radiation in an Atom Laser}
\author{B. M. Henson}
\affiliation{Laser Physics Centre, Research School of Physics and Engineering, The Australian National University,\\ Canberra, ACT 2601, Australia}
\author{Xuguang Yue}
\affiliation{State Key Laboratory of Magnetic Resonance and Atomic and Molecular Physics,Wuhan Institute of Physics and Mathematics, Chinese Academy of Sciences,Wuhan 430071, China}
\author{S. S. Hodgman}
\affiliation{Laser Physics Centre, Research School of Physics and Engineering, The Australian National University,\\ Canberra, ACT 2601, Australia}
\author{D. K. Shin}
\affiliation{Laser Physics Centre, Research School of Physics and Engineering, The Australian National University,\\ Canberra, ACT 2601, Australia}
\author{L. A. Smirnov}
\affiliation{Institute for Applied Physics, Russian Academy of Sciences, Nizhny Novgorod, Russia}
\affiliation{Research Institute for Supercomputing, Lobachevsky State University of Nizhny Novgorod, Nizhny Novgorod, Russia}
\author{E. A. Ostrovskaya}
\affiliation{Nonlinear Physics Centre, Research School of Physics and Engineering, The Australian National University,\\ Canberra, ACT 2601, Australia}
\author{X. W. Guan}
\affiliation{State Key Laboratory of Magnetic Resonance and Atomic and Molecular Physics,Wuhan Institute of Physics and Mathematics, Chinese Academy of Sciences,Wuhan 430071, China}
\author{A. G. Truscott}
\affiliation{Laser Physics Centre, Research School of Physics and Engineering, The Australian National University,\\ Canberra, ACT 2601, Australia}
\email{andrew.truscott@anu.edu.au}
\date{\today}

\begin{abstract}
We develop a simple yet powerful technique to study Bogoliubov-Cherenkov radiation by producing a pulsed atom laser from a strongly confined Bose-Einstein condensate.
Such radiation results when the atom laser pulse falls past a Bose-Einstein condensate at high-hypersonic speeds, modifying the spatial profile to display a characteristic twin jet structure and a complicated interference pattern. The experimental observations are in excellent agreement with mean-field numerical simulations and an analytic theory. Due to the highly hypersonic regime reached in our experiment, this system offers a highly controllable platform for future studies of condensed-matter analogs of quantum electrodynamics at ultrarelativistic speeds.
\end{abstract}



\maketitle

\section{Introduction}
Superfluidity is an important concept that was first rigorously defined by Landau \cite{khalatnikov2000introduction}. Under this definition, a weak repulsive potential (impurity) can travel through a superfluid without experiencing a friction force, so long as its motion remains below some critical velocity. Above the critical velocity it becomes energetically favorable for the motion of the impurity to cause perturbations in the fluid and superfluidity is lost. The resulting friction force felt by the moving impurity is due to the emission of elementary excitations in the fluid, which correspond to the quantum fluid analogue of Cherenkov radiation.

Many physical systems demonstrate Cherenkov emission, for example a charged particle traveling relativistically through a dielectric medium \cite{landau1984electrodynamics} or an object moving through a fluid at supersonic speed.
The details of the radiated field that is observed, however, depend on the excitation spectrum of the specific medium.
For a superfluid, which is described by Bogoliubov excitations, one expects to observe Bogoliubov-Cherenkov radiation (BCR) for super-sonic motion, which has previously been shown to give rise to standing wave patterns and a Cherenkov cone \cite{Carusotto2013}. 

Due to its macroscopic quantum properties \cite{RevModPhys.71.463} a gaseous Bose-Einstein condensate (BEC) can be considered a quantum fluid. 
As such, motion of objects through a BEC provide an ideal platform to study the superfluid behavior of a quantum fluid. 
To date, a number of experiments have probed superfluidity by using an optical potential as a moving object, and studying the disturbances produced in the expanded BEC cloud.
In these experiments, the critical velocity of a condensate has been investigated using calorimetry \cite{PhysRevLett.83.2502} as well as by measuring vortex production \cite{PhysRevLett.85.2228, PhysRevLett.104.160401}. The drag force an object experiences has been investigated theoretically \cite{PhysRevLett.69.1644, PhysRevLett.82.5186}.
Dispersive shock waves (DSW) have also been experimentally observed \cite{HoeferAblowitzCoddingtonEtAl2006, MeppelinkKollerVogelsEtAl2009,Dutton663} and theoretically modelled \cite{EL2006192, HoeferAblowitzCoddingtonEtAl2006}. These and related phenomena are also accessible in polaritonic condensates \cite{RevModPhys.85.299}.

In this paper we show, surprisingly, that BCR is present when atoms are simply outcoupled from a condensate in the form of an atom laser \cite{PhysRevLett.78.582,BlochHaenschEsslinger1999,GerbierBouyerAspect2001}. Unlike the only previous observation of BCR in a quantum fluid  \cite{CarusottoHuCollinsEtAl2006}, which involved the motion of a constant velocity laser beam (impurity) through a condensate, in our experiment the impurity (condensate) effectively accelerates during its passage through the quantum fluid.  As such the radiation pattern is distinctly different than that previously observed, and allows us to detect a Cherenkov bubble as the impurity exits the quantum fluid. We develop a comprehensive analytic theory of the effect and demonstrate remarkable agreement of the experiment with mean-field numerical simulations, thus providing complete quantitative description of the BCR dynamics. The size of the impurity and the speed of sound in this system can be well controlled by the outcoupling process, thus enabling us to test a wide range of flow speeds, including the high-hypersonic range. This feature, together with the impurity's acceleration, makes our system potentially useful as an analog testbed for a variety of exotic effects predicted to occur at relativistic and ultra-relativistic speeds in different phsyical systems.

\begin{figure*}[t] 
\centering
\includegraphics[width=17.8cm]{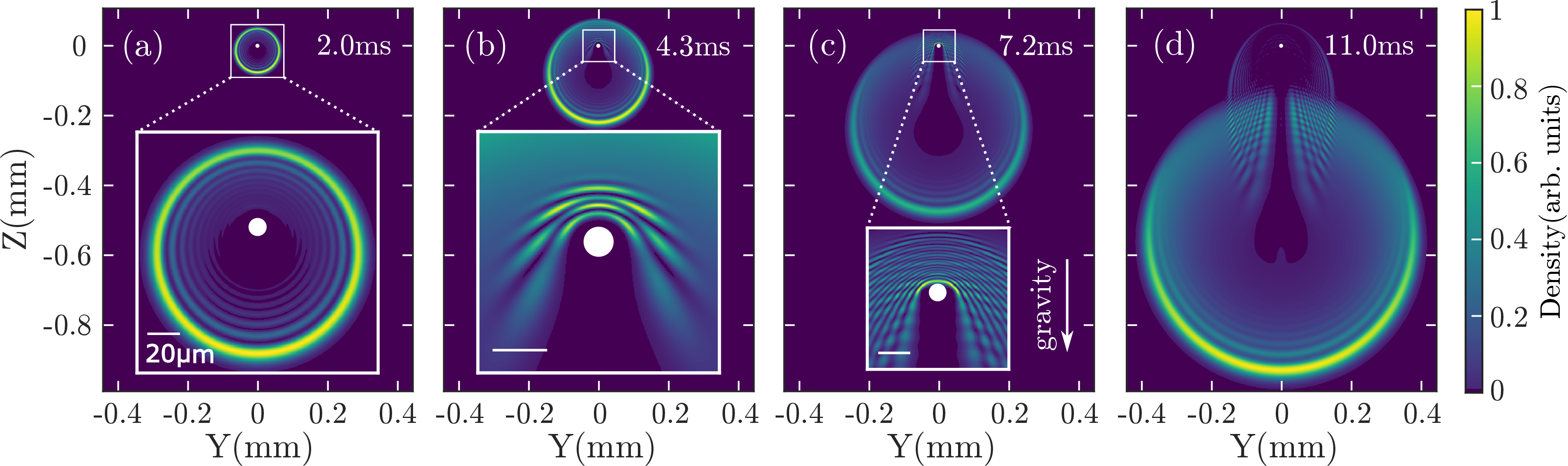}
\caption{
(Color online) (video available) Numerical simulations showing the formation of BCR in the atom laser beam. An RF pulse uniformly transfers a small number of trapped BEC atoms into the untrapped state, which then expand rapidly (a) generating a DSW as in \cite{HoeferAblowitzCoddingtonEtAl2006}. The atoms also begin to fall under gravity, causing the BEC to pass through them, generating a BCR bow wave in front of BEC as seen in (b). Once the BEC reaches  the edge of the atom laser (c) a complicated interference pattern generated by BCR is observed. Finally the BEC exits the atom laser yielding a Cherenkov bubble which turns into two characteristic jets in the far field (d). Each image corresponds to a time after RF outcoupling of: (a) 2.0~ms, (b) 4.3~ms, (c) 7.2~ms and (d) 11.0~ms. The ($11$~\textmu{}m) BEC is indicated by the while dot in each image while the scale bar in the inset indicates $20$~\textmu{}m.}
\label{fig:TheoryTimeSeries} 
\end{figure*}

In our experiment, we demonstrate excitation of BCR in a dilute atom laser comprised of metastable helium (He*) atoms. When only a small fraction of atoms are outcoupled, the repulsive mean-field potential from the trapped condensate causes the atom laser to expand rapidly. 
The expanded atom laser forms a dilute, inhomogeneous and anisotropic fluid, which then falls under gravity past the trapped condensate. 
This scenario is equivalent, via the Galilean transformation, to an accelerating impurity (the BEC) passing through a quantum fluid (the atom laser), in the rest frame of the atom laser. 
Previous studies of atom laser beams have revealed caustics \cite{RiouGuerinCoqEtAl2006}, interference fringes \cite{Dall2007}, four-wave mixing artifacts \cite{DallByronTruscottEtAl2009,RuGway2011}, and Heisenberg limited performance \cite{JeppesenDugueDennisEtAl2008} in the spatial profile.
The phenomena detailed here has not been observed in the multitude of atom laser experiments as it requires high confinement of the initial BEC and the summation of many experimental realizations.

Here, we show that BCR modifies the wavefunction of a pulsed atom laser in a striking manner in both momentum and real space. BCR occurs in our system because the relative velocity of the condensate passing through the atom laser exceeds the local speed of sound. As a result a standing wave forms in the atom laser beam, due to the reflection of incoming waves off the BEC, reminiscent of a bow wave that forms in front of a ship.  Since the atom laser has an inhomogeneous density profile, the local speed of sound is temporally dependent, resulting in bow waves of different wavelengths and thus an atom laser beam with a complicated cross hatching pattern (see Fig.~\ref{fig:TheoryTimeSeries}). 
We explain the observed pattern qualitatively using analytic scattering theory, while a two-dimensional (2D) simulation using the Gross-Pitaevskii (GP) equation yields an excellent agreement with the experimental observations.

\begin{figure*}[t] 
\centering
\includegraphics[width=17.8cm]{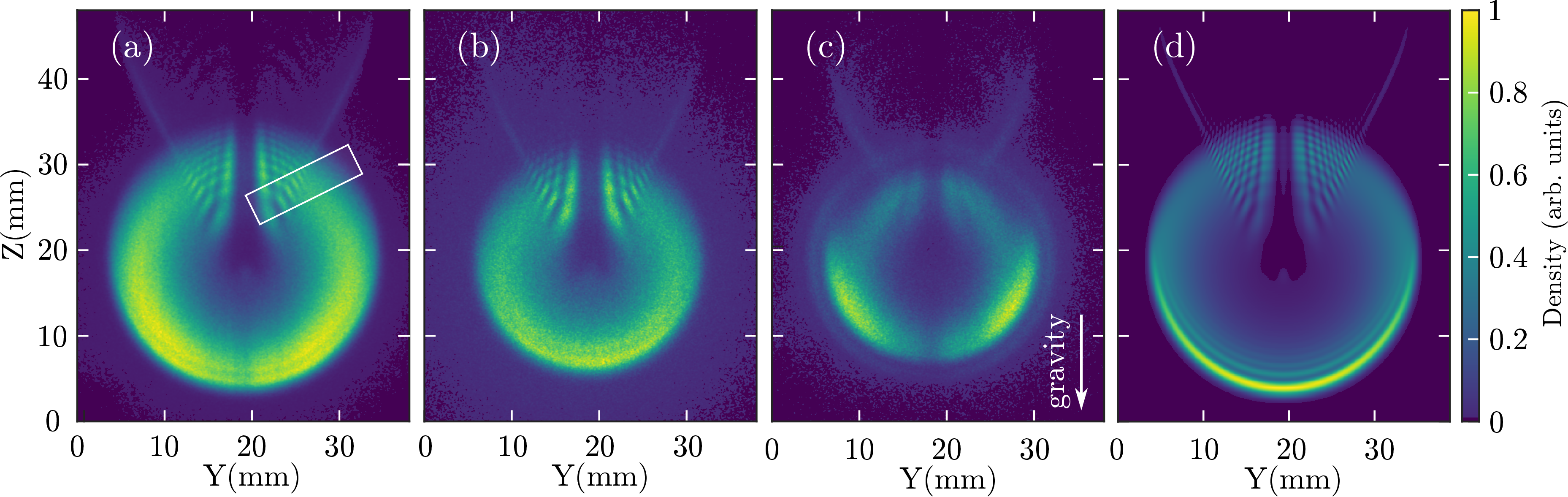}
\caption{ 
(Color online) Experimental observation of BCR resulting from the transit of a BEC  through an atom laser.
(a) For high condensate number ($N_0=7.0(3)\times10^{\textrm{5}}$), tight magnetic trap ($\{\omega_r,\omega_x\}\approx2\pi\{550,50\}\textrm{Hz}$) and weak outcoupling ($N_{\textrm{AL}}/N_{\textrm{0}}=3.5(3)\%$), the remnants of a bow wave is visible as twin jets extending above the atom laser, while strong interference fringing is visible in the main body. White box shows region integrated for comparison of fringe spacing with theoretical model in Fig. ~\ref{fig:data_BCR_theory}.
(b) With lower atom number ($N_0=3.7(2)\times10^{\textrm{5}}$) the results are qualitatively the same, albeit with a smaller spatial extent due to the reduced mean field repulsion.
(c) Upon dramatically increasing the outcoupling fraction ($N_0=1.7(3)\times10^{\textrm{5}}$, $N_{\textrm{AL}}/N_{0}=67(2)\%$) the fringes and jets become washed out, a faint ring is visible indicating the DSW survives to the far field and the atom laser profile is distorted. 
(d) The results of a two-dimensional numerical simulation of (a) extended to the detector position  shows strong qualitative agreement with (a). 
For an experimental atom laser profile at low confinement see Fig.~\ref{fig:low_trap_freq}. All experimental sub-figures are the summation of thousands of experimental realizations integrated along the weak (x) axis of the trap. 
The density in each image is normalized to the maximum pixel value.}
\label{fig:RFALShocks}
\end{figure*}

\section{Experimental setup}
The atom laser beam in our experiments is produced by illuminating a magnetically trapped BEC of metastable helium (He*) atoms, in the long lived $2^3S_1$ state \cite{Hodgman2009a}, with a  pulse  of radio frequency (RF) radiation.
The RF pulse, resonant with the Zeeman splitting between the $m_J{=}+1$ (trapped) and $m_J{=}0$ (untrapped) internal states at ${\sim} 700$~kHz, transfers atoms from the cigar-shaped trap, $\{\omega_r,\omega_x\}\approx2\pi\times\{550, 50\}$~Hz, into the atom laser beam. 
The short pulse duration (${\sim}10$~\textmu{}s) leads to a large broadening in the frequency spectrum, resulting in a nearly uniform outcoupling into the atom laser beam despite the inhomogeneous magnetic field of the trap.
Atoms then expand under the (cigar-shaped) repulsive mean field potential of the trapped condensate to form a thin disk with a (far field) spatial ratio of approximately ten to one. The subsequent passage of the trapped BEC through the atom laser can thus be treated as a 2D problem.

The atom laser then falls under gravity a distance of ${\sim} 850$~mm (time of flight ${\sim} 416$~ms) where the atoms are imaged in the far-field with full three dimensional resolution using an $80$~mm diameter multi-channel plate and delay line detector (DLD) \cite{Manning:10}. 
The large internal energy of the $2^3S_1$ state of He* (${\sim} 20$~eV) allows the DLD to reconstruct individual atoms at a spatial resolution of ${\sim} 120$~\textmu{}m, a temporal resolution of ${\sim} 3$~\textmu{}s (see Appendix \ref{apx_det_res}) and a quantum efficiency of ${\sim} 10\%$.

\section{Theory}
\subsection{Overview}
The dynamics of the atom laser outcoupling process are most clearly illustrated using the simulations of our experiment by means of the 2D mean-field Gross-Pitaevskii (GP) model. Fig.~\ref{fig:TheoryTimeSeries} shows a time series of the simulation that demonstrates this process in the near-field for the early stages of dynamics, which cannot be probed directly in our experiment. The outcoupling pulse is applied at $t{=}0$~ms, and the magnetically insensitive $m_J{=}0$ atoms undergo a rapid expansion due to the repulsive mean field potential from the trapped BEC (see Fig.~\ref{fig:TheoryTimeSeries}(a)). 

The expansion dynamics of the atom laser is therefore analogous to the DSW experiment reported in \cite{HoeferAblowitzCoddingtonEtAl2006} where a hard-wall optical potential is introduced non-adiabatically into a quantum fluid. Here it is rather the magnetic potential balancing the repulsive mean field potential that is removed for outcoupled atoms. Thus the trapped BEC serves as the barrier and the outcoupled atom laser as the quantum fluid.

As a result, we observe density modulations resembling the DSW from \cite{HoeferAblowitzCoddingtonEtAl2006} in the early dynamics of the pulsed atom laser, seen as the circular fringes in Fig.~\ref{fig:TheoryTimeSeries}(a). 
The atom laser falls past the trapped condensate at an accelerating speed that quickly ($t\sim1ms$) becomes supersonic. By $t{=}4.3$~ms (see Fig.~\ref{fig:TheoryTimeSeries}(b)) a significant portion has already passed the BEC. 
The resulting high-hypersonic flow yields a series of wavefronts and a conical shadow region begins to develop, reminiscent of BCR \cite{CarusottoHuCollinsEtAl2006}. 
While the relative velocity of the BEC passing through the atom laser due to gravity is modest, the low density in the atom laser means that the passage of the barrier is in the high-hypersonic regime, with a Mach number of $M{\sim}10 \textrm{-} 100$ (see  Fig.~\ref{fig:mach_dyn}).
By $t{=}7.2$~ms (Fig.~\ref{fig:TheoryTimeSeries}(c)), when the BEC is almost at the edge of the atom laser, several bow waves have been generated with differing wavelengths due to the time varying local speed of sound and flow velocity, and the characteristic cross-hatching pattern is formed. As the BEC exits the atom laser ($t{=}11$~ms, Fig.~\ref{fig:TheoryTimeSeries}(d)), the bow waves separate from the atom laser completely forming a Cherenkov bubble in the near field that appear as two characteristic trailing `jets' in the far field on our detector.

\subsection{2D Gross Pitaeveski Simulations}
The experimental process for generating BCR in an atom laser beam is well described by the following coupled GP equations:
\begin{align}
i\hbar\frac{\partial}{\partial t}\Psi_{c}\left(\mathbf{r},t\right) = &\left[-\frac{\hbar^{2}}{2m}\Delta+V_{\textrm{trap}}\left(\mathbf{r}\right)+U_{cc}\left|\Psi_{c}\right|^{2}\right.\nonumber\\
&+\left.U_{ac}\left|\Psi_{a}\right|^{2}\right]\psi_{c}+\hbar\Omega\Psi_{a},\label{eq:CGPE1}\\
i\hbar\frac{\partial}{\partial t}\Psi_{a}\left(\mathbf{r},t\right) = & \left[-\frac{\hbar^{2}}{2m}\Delta + mgz+U_{aa}\left|\Psi_{a}\right|^{2}\right.\nonumber\\
&+\left.U_{ac}\left|\Psi_{c}\right|^{2}\right]\psi_{a}+\hbar\Omega\Psi_{c},\label{eq:CGPE2}
\end{align}
where $\Psi_{c(a)}$ is the condensate (atom laser) wave function,   $V_{\textrm{trap}}\left(\mathbf{r}\right)=\frac{1}{2}m\left(\omega_{x}^{2}x^{2}+\omega_{y}^{2}y^{2}+\omega_{z}^{2}z^{2}\right)$ is the harmonic trap potential with the trapping frequencies $\omega_{z}$, $\omega_{y}$ and $\omega_{z}$,  $U_{cc}$, $U_{aa}$ and $U_{ac}$ are the interaction strengths between condensate atoms, atoms in the atom laser, and between atoms in the condensate and atoms in the atom laser, respectively.  $\Omega$ is the Rabi frequency of the RF coupling, and $g$ is the acceleration due to gravity along the $-z$  direction. The experimental trapping frequencies are $\omega_x = 2\pi\times 50~\text{Hz}$, $\omega_{y}=\omega_{z}=\omega_{r}=2\pi\times550~\text{Hz}$, and the interaction strengths are $U_{cc}=U_{ac}=4\pi\hbar^{2}a_{1}/m$, $U_{aa}=4\pi\hbar^{2}a_{0}/m$ with the $s$-wave scattering lengths $a_{1}=7.51$~nm \cite{PhysRevLett.96.023203} and $a_{0}=5.56$~nm \cite{PhysRevA.64.042710}.

The experiment can be separated into two stages, namely the rapid RF outcoupling stage followed by a stage during which the condensate wavefunction can be considered constant. The first stage is very quick and lasts for hundreds of micro-seconds to milli-seconds, while the second one lasts for up to a half second. In the first order approximation, at the end of the first stage, the atom laser wave function equals the ground state wave function of the condensate in the trap. The latter can be found numerically by solving Eq. (\ref{eq:CGPE1}) without the RF coupling term (using, e.g., the imaginary-time evolution method). After that, the two atomic clouds, i.e., the trapped BEC and the atom laser, evolve separately, with each one providing an effective potential for the other. 
However, in the experiment, the outcoupled atoms constitute only a few percent of the condensate, therefore the back action of the atom laser on the condensate can be safely ignored. This feature has also been confirmed numerically.

Based on the above discussion, after turning off the RF coupling, the condensate wave function can be treated as a constant, $\Psi_{c} (\mathbf{r},t){\simeq}\sqrt{N_\textrm{BEC}/N_{0}} \Psi_{g}(\mathbf{r})$, where $\Psi_{g}(\mathbf{r})$ is the ground state wave function of the condensate in the trap and $N_\textrm{BEC}/N_{0}$ is the fraction of atoms that remain in the condensate after the atom laser ($N_\textrm{BEC}=N_{0}-N_\textrm{AL}$). The evolution of the atom laser can then be described by a one-component GP model:
\begin{align}
  i\hbar\frac{\partial\Psi}{\partial t}=&\left(-\frac{\hbar^{2}}{2m} \Delta+mgz+U_{aa}\left|\Psi\right|^{2}\right.+\left.U_{ac}\left| \Psi_{g}\right|^{2}\right)\Psi,\label{eq:al}
\end{align}
where the index $a$ has been omitted for simplicity. Thus, the dynamics of the atom laser is determined by the gravity and the potential barrier provided by the condensate. The initial condition is $\Psi(\mathbf{r}, t=0)=\sqrt{N_\textrm{AL}/N_{0}}\Psi_{g}(\mathbf{r})e^{i\phi}$, where $N_\textrm{AL}$ and $N_{0}$ is the number of atoms in the atom laser and initial (before atom laser) condensate, respectively. Since the RF out-coupling process is fast and coherent, we can take the global phase as $\phi=0$.

To solve Eq. (\ref{eq:al}) numerically, we transform it into a dimensionless form. To this end, we introduce the following scalings:
\begin{align*}
\tilde{t}=&\omega_{r}t,\quad\tilde{\mathbf{r}}=\mathbf{r}/l_{r},
\quad l_{r}=\sqrt{\frac{\hbar}{m\omega_{r}}},\\
\quad\tilde{\Psi}=&\frac{l_{r}^{3/2}}{\sqrt{N_\textrm{AL}}}\Psi,\quad
\tilde{\Psi}_{g}=\frac{l_{r}^{3/2}}{\sqrt{N_{BEC}}}\Psi_{g}.
\end{align*}
Then Eq.(\ref{eq:al}) can be re-written in the following dimensionless form:
\begin{equation}
i\frac{\partial\Psi}{\partial t}=\left(-\frac{1}{2}\Delta
-g^{\textrm{eff}}z+U_{aa}^{\textrm{eff}}\left|\Psi\right|^{2}
+U_{ac}^{\textrm{eff}}\left|\Psi_{g}\right|^{2}\right)
\Psi\,\label{eq:aldimless}
\end{equation}
where
\[
g^{\textrm{eff}}=\frac{g}{l_{r}\omega_{r}^{2}},\quad
U_{aa}^{\textrm{eff}}=\frac{4\pi N_\textrm{AL}a_{0}}{l_{r}},\quad
U_{ac}^{\textrm{eff}}=\frac{4\pi N_\textrm{BEC}a_{1}}{l_{r}}.
\]
In the above equation we have dropped the tildes from the relevant variables.

Furthermore, since the longitudinal trap frequency ($\omega_{x}$) is much smaller than the transverse ones ($\omega_{r}$), the atom laser expands much slower in the longitudinal direction. We can therefore ignore the expansion in the longitudinal direction and integrate it out. This can be done by separating the variables and factorizing the wave function as follows:
\begin{equation}
\Psi(\mathbf{r},t)=\psi(y,z,t)\phi(x)\exp\left(-i\lambda t/2\right)\label{eq:sov}
\end{equation}
where $\phi(x)=(\lambda/\pi)^{1/4}\exp(-\lambda x^{2}/2)$, and $\lambda=\omega_{x}/\omega_{r}$. The density of the condensate can be approximated by Thomas-Fermi (TF) profile, that is $\left|\Psi_{g}\right|^{2}=n_{0}\left(1-\frac{x^{2}}{R_{x}^{2}}-\frac{y^{2}+z^{2}} {R_{r}^{2}}\right)$, where $n_{0}$ is the peak density, and the TF radius is $R_{x}$ and $R_{r}$ in the transverse and longitudinal directions, respectively. Substituting Eq. (\ref{eq:sov}) into (\ref{eq:aldimless}), multiplying both sides by $\phi(x)$ and integrating over $x$, we obtain the 2D evolution equation for the atom laser wave function:
\begin{align}
i\frac{\partial\psi(y,z,t)}{\partial t}&=\left[-\frac{1}{2}\left(\frac{\partial^{2}}
{\partial y^{2}}+\frac{\partial^{2}}{\partial z^{2}}\right)+g^{\textrm{eff}}z\right.\nonumber\\
&+\left.\frac{U_{aa}^{\textrm{eff}}}{\sqrt{2\pi/\lambda}}\left|\psi\right|^{2}
+U_{ac}^{\textrm{eff}}\left|\psi_{g}\right|^{2}\right]\psi(y,z,t),\label{eq:alfinal}
\end{align}
where we have dropped a constant term. 

The interaction between atoms in the atom laser is much weaker than that within the condensate. This is due to a combination of effects. First of all, as the scattering length is of the same order, the interaction strength is proportional to the atom number, $U_{aa}^{\textrm{eff}}/U_{ac}^{\textrm{eff}}{\sim} N_\textrm{AL}/N_{BEC}$. In our experiments, this ratio is about $1-4\%$. Secondly, the interaction strength in the atom laser is further reduced by a factor of $\sqrt{2\pi/\lambda}$, which is about $8$ for the experimental value of $\lambda=1/11$. These two effects reduce the interaction strength between the atoms in the atom laser by at least two order of magnitude compared to that within the condensate.

Eq. (\ref{eq:alfinal}) is readily solved numerically \cite{Bao:2003hh, Vudragovic:2012jz,YoungS:2016fh} by the so-called Time-Splitting Spectral method \cite{Bao:2003hh} with the FFTW package \cite{FFTW05}. The results are shown in Fig.~\ref{fig:TheoryTimeSeries}. The simulation also predicts the formation of vortices (see Appendix~\ref{apx_gp_sim}).

\subsection{Analytic model}

\begin{figure*}[th]
\centering
\includegraphics[width=\linewidth]{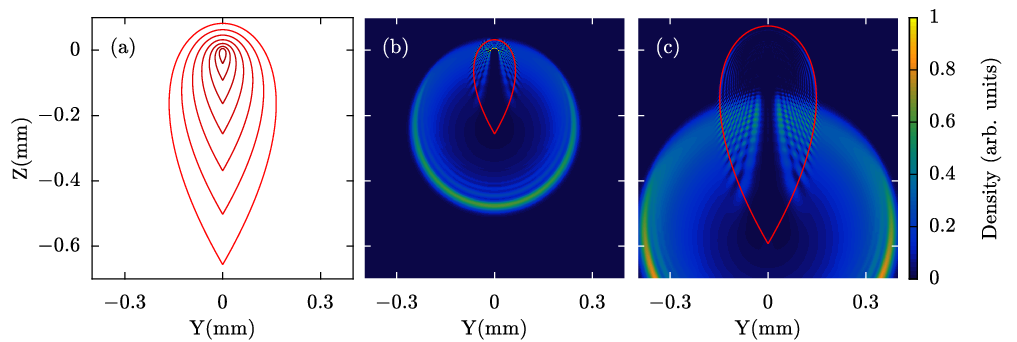}
\caption{(Color online) (a) Theoretical position of the caustics calculated using Eqs.~\eqref{eq:characteristics} for different moments of time: $\omega_{r}t=10:5:40$ (in dimensional variables). (b), (c) Atom laser density profiles at the times $\omega_{r}t=25$ and $\omega_{r}t=38$, respectively. Spatial distributions of the density were obtained by direct numerical simulation performed immediately within the framework of the one-component 2D GP equation. The solid red curves show the corresponding spatio-temporal caustic lines, which were calculated using Eqs.~\eqref{eq:characteristics}. Here, $\omega_{r}=2{\pi}\cdot550{\,}\textrm{Hz}$ and $\textsl{g}=g\bigl/\omega_{r}l_{r}\bigr.=0.383$. The analytic model provides good qualitative comparison with the typical numerical simulations of the experimentally observed effect.}
\label{fig:caustic2}
\end{figure*}

An analytical model can be used to describe the underlying standing-wave patterns and the trailing jets of the atom laser scattering off the BEC. 
In constructing this theoretical description, we have used the following considerations. (i) The experimental dynamics are sufficiently well described by the 2D one-component GP equation. 
(ii) Since the characteristic transverse size of the condensate is significantly smaller than the typical value of the local healing length in the atom laser, one may neglect atom-atom interactions between atoms in the atom laser cloud. 
(iii) We approximate the potential barrier (caused by the BEC) by an impenetrable obstacle of cylindrical shape, since the density of the BEC is much larger than that of the atom laser, $n_{\textrm{BEC}} \gg n_{\textrm{AL}}$.
Within these assumptions, we derive an equation for the curve near which the instantaneous maxima of the particle density in the scattered component of the atom laser wave function should be located. Agreement between our experiment and this analytic model confirms the experimental features in the atom laser beam can be interpreted as the result of interference of incident and scattered waves from a hard-wall potential barrier.

As discussed above, by virtue of the configuration features of the initial BEC cloud, the expansion of the wave function $\Psi_{a}\left(\vec{r},t\right)$ of the outcoupled atoms in the $y$, $z$ plane is much more rapid than that along the $x$ coordinate. Moreover, if the number of atoms in the atom laser, $N_\textrm{AL}$, is much smaller than the initial number of particles in the system, $N_{0}$, then the atom laser does not affect the behavior of the trapped condensate, which plays the role of a potential barrier for $\Psi_{a}\left(\vec{r},t\right)$. According to the previous Section, the dynamics of the atom laser beam is well described by a 2D one-component GP equation. 
In this equation, one can, in principle, get rid of the gravitational potential term proportional to $\textsl{g}z$ by introducing a coordinate system moving in the negative direction of the $z$ axis with constant acceleration $\textsl{g}$. Thus, the problem of the atom laser flow past an obstacle, can be interpreted as generation of excitations by a uniformly accelerated object in an inhomogeneous ultracold degenerate Bose gas. The mechanism for the formation of a cross-hatching pattern in the upper part of the atom laser beam therefore has the character of Cherenkov radiation of density waves with the Bogoliubov spectrum~(e.g., see\,\cite{CarusottoHuCollinsEtAl2006, PhysRevA.75.033619, 0953-4075-41-16-165301, CarreteroGonzález2007361, Mironov2010}).

Experimental measurements and estimates show that the relative velocity of the effective potential barrier greatly exceeds the local speed of sound in the atom laser cloud. In turn, the size of the obstacle and the characteristic spatial scale of the observed wave pattern are much smaller than the typical healing length. In addition, as seen both in experiment and in the numerical simulations, the density of the atom laser cloud in the region of interest decreases with time, and becomes strongly rarefied. Hence, it is reasonable to neglect a nonlinear term in the 2D one-component GP equation. 

Thus, for the further analytical description of the processes, we will use a 2D linear Schr\"{o}dinger equation, which has the following dimensionless form similar to~\eqref{eq:alfinal}:
\begin{equation}\label{eq:SchrodingerEq}
i\frac{\partial\psi}{\partial{\tau}}=-\frac{1}{2}\left(\frac{\partial^{2}\psi}{\partial{y}^{2}}+\frac{\partial^{2}\psi}{\partial{z}^{2}}\right)+V\left(y,z\right)\psi+\textsl{g}z\psi,
\end{equation}
where $V\left(y,z\right)$ is the effective potential barrier formed by the stationary density distribution of the trapped BEC. The characteristic transverse scale $R$ of this cigar-shaped cloud does not exceed a few dimensionless units. Here, for convenience, we introduced also the time $\tau=t-\delta{t}$ with the coordinate origin shifted by a constant value $\delta{t}$. This value $\delta{t}$ characterizes a relatively short time interval, during which the outcoupled atoms are displaced from the localization area of the trapped condensate confined by a magnetic field.

The solution to Eq.~\eqref{eq:SchrodingerEq} will be sought in the form of a superposition of two fields:
\begin{equation}\label{eq:PsiDecomposition}
\psi\left(y,z,\tau\right)=\psi_{i}\left(y,z,\tau\right)+\psi_{r}\left(y,z,\tau\right).
\end{equation}
The first term, $\psi_{i}\left(y,z,\tau\right)$, describes the laser beam incident on an obstacle. The second term, $\psi_{r}\left(y,z,\tau\right)$, is the wave packet formed by the atom laser atoms reflected by the effective potential barrier.

The wave function $\psi_{i}\left(y,z,\tau\right)$ in Eq.~\eqref{eq:PsiDecomposition} satisfies the equation for a free atom laser beam in the absence of an obstacle
\begin{equation}\label{eq:SchrodingerEq-i1}
i\frac{\partial\psi_{i}}{\partial{\tau}}=-\frac{1}{2}\left(\frac{\partial^{2}\psi_{i}}{\partial{y}^{2}}+\frac{\partial^{2}\psi_{i}}{\partial{z}^{2}}\right)+\textsl{g}z\psi_{i}
\end{equation}
over the entire space. If we assume that for $\tau=0$, i.e., at the instant $t=\delta{t}$ when the RF outcoupling is completed, the initial distribution of the atom laser cloud is known, 
\begin{equation}\label{eq:psi-i_init}
\psi_{i}\left(y,z,\tau=0\right)=w\left(y,z\right),
\end{equation}
then it is easy to find a general solution for Eq.~\eqref{eq:SchrodingerEq-i1}. To do this, we introduce the accelerating coordinate system  $y$,~$\varsigma=z+\textsl{g}t^{2}/2$: 
\begin{equation}\label{eq:psi-i}
\psi_{i}\left(y,z,\tau\right)=\phi\left(y,\varsigma,\tau\right)e^{-i\textsl{g}{\tau}z-\left.i\textsl{g}^{2}{\tau}^{3}\!\right/6},
\end{equation}
and therefore exclude the gravitational potential term from Eq.~\eqref{eq:SchrodingerEq-i1}. Then for $\phi\left(y,\varsigma,\tau\right)$ we arrive at an equation similar to the Schr\"{o}dinger equation for a free electron. Using the Green's function of this equation 
\begin{equation}\label{eq:GreenFunction}
G\left(y-\tilde{y},\varsigma-\tilde{\varsigma},\tau\right)=\frac{1}{2i\pi\tau}e^{i\left.\left(\left(y-\tilde{y}\right)^{2}+\left(\varsigma-\tilde{\varsigma}\right)^{2}\right)\!\right/{2\tau}},
\end{equation}
we find: 
\begin{equation}\label{eq:phi}
\!\!\!\phi\!\left(y,\varsigma,\tau\right)\!=\!\frac{1}{2i\pi\tau}\!\iint\!{e^{i\left.\left(\left(y-\tilde{y}\right)^{2}+\left(\varsigma-\tilde{\varsigma}\right)^{2}\right)\!\right/{2\tau}}w\!\left(\tilde{y},\tilde{\varsigma}\right)\textsl{d}\tilde{y}\textsl{d}\tilde{\varsigma}}.\!\!
\end{equation}

We then take into account that the spatial scale of the distribution $w\left(y,z\right)$ is comparable to the transverse size $R$ of the trapped BEC cloud. Then for $\tau\gg{2R^{2}}$ the formula~\eqref{eq:phi} can be further simplified, and at a sufficiently long time $\tau$, the atom laser wavefunction $\psi_{i}\left(y,z,\tau\right)$ is well described by the expression:
\begin{equation}\label{eq:psi-i_f}
\psi_{i}\!\left(y,z,\tau\right)\!=\!\frac{1}{i\tau}\hat{w}\biggl(\frac{y}{\tau},\!\frac{z}{\tau}+\frac{\textsl{g}\tau}{2}\biggr)e^{i\!\left.\left(y^{2}+z^{2}\right)\!\right/2\tau-\left.i\textsl{g}{\tau}z^{\!\!\!\!\phantom{2}}\right/2-\left.ig^{2}\tau^{3\!}\right/24},
\end{equation}
where $\hat{w}\left(\varkappa_{1},\varkappa_{2}\right)$ is the Fourier transform of the initial distribution $\psi_{i}\left(y,z,0\right)=w\left(y,z\right)$:
\begin{equation}\label{eq:fourier}
\hat{w}\left(\varkappa_{1},\varkappa_{2}\right)=\frac{1}{2\pi}\!\iint\!{w\left(\tilde{y},\tilde{\varsigma}\right)e^{-i\varkappa_{1}\tilde{y}-i\varkappa_{2}\tilde{\varsigma}}\textsl{d}\tilde{y}\textsl{d}\tilde{\varsigma}}.
\end{equation}
Equation~\eqref{eq:psi-i_f} clearly shows that, for large $\tau$, the density $\left|\psi_{i}\!\left(y,z,\tau\right)\right|^{2}$ of particles in the wave packet $\psi_{i}\!\left(y,z,\tau\right)$ incident on an obstacle is smoothly varied depending on the coordinates $y$ and $z$ and remains almost constant over the length scales of the order of $R$. It also follows directly from Eq.~\eqref{eq:psi-i_f} that, for $\tau{\gg}2R^{2}$, the velocity field arising on a cylindrical surface of radius $R$, which is the boundary of the effective potential barrier, is mainly determined by the term $-i\textsl{g}{\tau}z\bigl/2\bigr.$ in the phase of the complex function $\psi_{i}\!\left(y,z,\tau\right)$ when the atom laser falls onto that surface.

As discussed above, $N_{0}/N_\textrm{AL}{\gg}1$, and therefore the effective potential barrier $V\!\left(x,y\right)$ in Eq.~\eqref{eq:SchrodingerEq} can be approximated by a hard-wall obstacle that is impenetrable to the atom laser and has the cylindrically symmetric form with the radius~$R$. As a result, the wave function $\psi_{r}\left(y,z,\tau\right)$ of the reflected wave packet in the region $\sqrt{y^{2}\!+\!z^{2}}\!>\!R$ obeys the following linear Schr\"{o}dinger equation:
\begin{equation}\label{eq:SchrodingerEq-r}
i\frac{\partial\psi_{r}}{\partial{\tau}}=-\frac{1}{2}\left(\frac{\partial^{2}\psi_{r}}{\partial{y}^{2}}+\frac{\partial^{2}\psi_{r}}{\partial{z}^{2}}\right)+gz\psi_{r}.
\end{equation}
By using the Madelung transformation,
\begin{equation}\label{eq:psi-r_a-varphi}
\psi_{r}\left(y,z,\tau\right)=a\left(y,z,\tau\right)\exp\left(i\varphi\left(y,z,\tau\right)\right),
\end{equation}
we can write down the equations for the amplitude $a\left(y,z,t\right)$ and phase $\varphi\left(y,z,t\right)$ of the reflected wavepacket:
\begin{gather}
\frac{\partial{a^{2}}}{\partial\tau}+\textrm{div}_{\!\perp}\!\left(a^{2}\nabla_{\!\perp}\varphi\right)=0,\label{eq:a-eq}{}\\{}
\frac{\partial\varphi}{\partial\tau}+\frac{1}{2}\left(\nabla_{\!\perp}\varphi\right)^{2}+gz=\frac{\Delta_{\!\perp}a}{2a}.\label{eq:phi-eq}
\end{gather}
Here, we have introduced the notation $\nabla_{\!\perp}$, $\textrm{div}_{\!\perp}$ and $\Delta_{\!\perp}$ for the gradient, divergence and Laplace operator taken with respect to the coordinates $y$ and $z$. In the case of the complete reflection of the dilute atom laser cloud by the impenetrable cylinder with the radius $R$, the boundary conditions on the top $(z>0)$ semi-circle $\sqrt{y^{2}+z^{2}}=R$ at each fixed moment of time $\tau_{0}$ can be set as follows:
\begin{gather}
\psi_{r}\left(R\sin{\vartheta},R\cos{\vartheta},\tau_{0}\right)\!=\!-\psi_{i}\left(R\sin{\vartheta},R\cos{\vartheta},\tau_{0}\right),\label{eq:psi-r_on-circle}{}\\{}
\nabla_{\!\perp}a\left(R\sin{\vartheta},R\cos{\vartheta},\tau_{0}\right)=0,\label{eq:nabla-a_on-circle}{}\\{}
\!\!\!\nabla_{\!\perp}\varphi\!\left(R\sin{\vartheta},R\cos{\vartheta},\tau_{0}\right)\!=\!\frac{g\tau_{0}}{2}\!\left(\sin{2\vartheta}\,\vec{y}_{0}\!+\!\cos{2\vartheta}\,\vec{z}_{0}\right)\!.\!\!\label{eq:nabla-phi_on-circle}
\end{gather}
where $\vartheta$ is the angle formed with the positive axis $z$: $-\pi/2<\vartheta<\pi/2$, and $\vec{y}_{0}$, $\vec{z}_{0}$ are the unit vectors along the directions $y$ and $z$, respectively. We note that the relations~\eqref{eq:psi-r_on-circle}, \eqref{eq:nabla-a_on-circle} and \eqref{eq:nabla-phi_on-circle} are written using Eq.~\eqref{eq:psi-i_f}, which holds for relatively large values of $\tau_{0}$ ($\tau_{0}\gg{R}^{2}$). This allows us to treat $\left|\psi_{i}\left(y,z,\tau\right)\right|$ as constant across the area of several $R$ in size and to only take into account the  term $-i\textsl{g}{\tau}z\bigl/2\bigr.$ in the phase of the complex function $\psi_{i}\left(y,z,\tau\right)$ when we calculate the velocity field created on the surface of the effective potential barrier by the incident wave packet $\psi_{i}\left(y,z,\tau\right)$.

The observed cross-hatching structure of spatial density distribution can be considered as a complex interference pattern of two wave functions $\psi_{i}\left(y,z,\tau\right)$ and $\psi_{r}\left(y,z,\tau\right)$ at the top part of the atom laser beam. The amplitude $\left|\psi_{i}\left(y,z,\tau\right)\right|$ of the incident wave packet varies smoothly over the spatial scales comparable to the size $R$ of the effective potential barrier. Hence, it is natural to assume that the amplitude $a\left(y,z,\tau\right)$ of the reflected field is a slowly varying function of the coordinates $y$ and $z$. Based on this assumption we can neglect the right-hand side in Eq.~\eqref{eq:phi-eq} and obtain the so-called eikonal equation:
\begin{equation}\label{eq:eikonal-eq}
\frac{\partial\varphi}{\partial\tau}+\frac{1}{2}\left(\frac{\partial{\varphi}}{\partial{y}}\right)^{2}+\frac{1}{2}\left(\frac{\partial{\varphi}}{\partial{z}}\right)^{2}+\textsl{g}z=0,
\end{equation}
This equation belongs to the class of Hamilton-Jacobi first-order differential equations~\cite{MaslovFedoryuk_1981}:
\begin{equation}\label{eq:H1}
\mathcal{H}\left(y,z,\tau,\frac{\partial{\varphi}}{\partial{y}},\frac{\partial{\varphi}}{\partial{z}},\frac{\partial{\varphi}}{\partial{\tau}}\right)=0
\end{equation}
or
\begin{equation}\label{eq:H2}
\mathcal{H}\bigl(\left\{q_{j}\right\},\left\{p_{j}\right\}\bigr)=0,\quad p_{j}=\frac{\partial\varphi}{\partial{q_{j}}},\quad j=1,2,3.
\end{equation}
Here by $\left\{q_{j}\right\}$ and $\left\{p_{j}\right\}$ we denote the set of the generalised position variables $y$, $z$ and $\tau$ and the set of the conjugate momentum variables $p_{y}$, $p_{z}$ and $p_{\tau}$, respectively.

The relation:
\begin{equation}
\textsl{d}\mathcal{H}=\sum_{j=1}^{3}\left(\frac{\partial\mathcal{H}}{\partial{q_{j}}}\textsl{d}{q_{j}}+\frac{\partial\mathcal{H}}{\partial{p_{j}}}\textsl{d}{p_{j}}\right)=0.
\end{equation}
should be satisfied on a hypersurface $\mathcal{H}\bigl(\left\{p_{j}\right\},\left\{q_{j}\right\}\bigr)\!=\!0$ in the phase space $\left\{p_{j}\right\}$,~$\left\{q_{j}\right\}$.
This occurs only if the following equalities are simultaneously satisfied~\cite{MaslovFedoryuk_1981}: 
\begin{equation}\label{eq:Heq}
\frac{\textsl{d}{q_{j}}}{\textsl{d}\xi}=\frac{\partial\mathcal{H}}{\partial{p_{j}}},\quad
\frac{\textsl{d}{p_{j}}}{\textsl{d}\xi}=-\frac{\partial\mathcal{H}}{\partial{q_{j}}},\quad
j=1,2,3,
\end{equation}
whee $\xi$ is an independent variable. The system of equations~\eqref{eq:Heq} expressed in the canonical Hamiltonian form represents a characteristic system of ordinary differential equations, and its solutions, $\left\{q_{j}\left(\xi\right)\right\}$ and $\left\{p_{j}\left(\xi\right)\right\}$, are the characteristics of the Hamilton-Jacobi equation~\eqref{eq:H1}.

Let us write down~\eqref{eq:Heq} explicitly by taking into account Eq.~\eqref{eq:eikonal-eq}:
\begin{subequations}\label{eq:eq-pj-qj}
\begin{eqnarray}
\frac{\textsl{d}y}{\textsl{d}\xi}&=&p_{y},\quad\frac{\textsl{d}p_{y}}{\textsl{d}\xi}=0,{}\\{}
\frac{\textsl{d}z}{\textsl{d}\xi}&=&p_{z},\quad\frac{\textsl{d}p_{z}}{\textsl{d}\xi}=-\textsl{g},{}\\{}
\frac{\textsl{d}\tau}{\textsl{d}\xi}&=&1_{\phantom{z}},\quad\frac{\textsl{d}p_{\tau}}{\textsl{d}\xi}=0.
\end{eqnarray}
\end{subequations}
From this equations, it is obvious that $\xi=\tau$, and
\begin{subequations}\label{eq:eq-yz-1}
\begin{gather}
y=y_{0}+p_{y}\left(\tau_{0}\right)\left(\tau-\tau_{0}\right),{}\\{}
z=z_{0}+p_{z}\left(\tau_{0}\right)\left(\tau-\tau_{0}\right)-\textsl{g}\left(\tau-\tau_{0}\right)^{2}\!\bigl/2\bigr..
\end{gather}
\end{subequations}
Here $y_{0}=R\sin{\vartheta}$, $z_{0}=R\cos{\vartheta}$ are the coordinates of a point on a semi-sphere, and $p_{y}\left(\tau_{0}\right)=\textsl{g}\tau_{0}\sin{2\vartheta}\bigl/2\bigr.$, $p_{z}\left(\tau_{0}\right)=\textsl{g}\tau_{0}\cos{2\vartheta}\bigl/2\bigr.$, as follows from Eq.~\eqref{eq:nabla-phi_on-circle} and the definition $\vec{p}\left(\tau\right)=\nabla_{\!\perp}\varphi\left(y\left(\tau\right),z\left(\tau\right),\tau\right)$. Finally, the characteristics of Eq.~\eqref{eq:eikonal-eq} are given by the expressions:
\begin{subequations}\label{eq:eq-yz-2}
\begin{gather}
y\left(\tau\right)=R\sin{\vartheta}+\textsl{g}\tau_{0}\sin{2\vartheta}\left(\tau-\tau_{0}\right)\!\bigl/2\bigr.,\label{eq:eq-yz-2-y}{}\\{}
\!\!z\!\left(\tau\right)\!=\!R\cos{\vartheta}\!+\!\textsl{g}\tau_{0}\cos{2\vartheta}\left(\tau\!-\!\tau_{0}\right)\!\bigl/2\bigr.\!-\!\textsl{g}\left(\tau\!-\!\tau_{0}\right)^{2}\!\bigl/2\bigr..\!\label{eq::eq-yz-2-z}
\end{gather}
\end{subequations}

The Eq.~\eqref{eq:a-eq} takes the form of a continuity equation. It allows us to determine behaviour of the density $a^{2}\!\left(y\!\left(\tau\right),z\!\left(\tau\right),\tau\right)$ of atoms in the reflected wave packet along the characteristics~\eqref{eq:eq-yz-2} of the eikonal equation~\eqref{eq:eikonal-eq}. To solve Eq.~\eqref{eq:a-eq}, we consider this equality on one of the curves $y\!\left(\tau\right)$, $z\!\left(\tau\right)$, $p_{y}\!\left(\tau\right)$, $p_{z}\!\left(\tau\right)$ in the phase space $y$, $z$, $p_{y}$, $p_{z}$ and re-write its left-hand side in the following form:
\begin{multline*}
\frac{\partial{a^{2}}}{\partial\tau}+\nabla_{\!\perp}{\varphi}\nabla_{\!\perp}{a^{2}}+a^{2}\textrm{div}_{\!\perp}\nabla_{\!\perp}{\varphi}=
\left(\frac{\partial{a^{2}}}{\partial\tau}+\vec{p}\left(\tau\right)\nabla_{\!\perp}{a^{2}}\right){}\\{}+a^{2}\textrm{div}_{\!\perp}\!\left(\vec{p}\bigl(\tau\right)\bigr)=\frac{\textsl{d}a^{2}}{\textsl{d}\tau}+a^{2}\textrm{div}_{\!\perp}\!\bigl(\vec{p}\left(\tau\right)\bigr).
\end{multline*}
As a result, we obtain~\cite{MaslovFedoryuk_1981}:
\begin{equation}\label{eq:continuity-eq}
\frac{\textsl{d}}{\textsl{d}\tau}\ln\bigl(a^{2}\left(\tau\right)\bigr)=-\,\textrm{div}_{\!\perp}\bigl(\vec{p}\left(\tau\right)\bigr).
\end{equation}
According to the Liouville theorem~\cite{MaslovFedoryuk_1981}:
\begin{equation}\label{eq:LiouvilleTheorem}
\textrm{div}_{\!\perp}\bigl(\vec{p}\left(\tau\right)\bigr)=\frac{\textsl{d}}{\textsl{d}t}\ln\bigl(\mathcal{J}\left(\tau,\tau_{0},\vartheta\right)\bigr),
\end{equation}
where $\mathcal{J}\left(\tau,\tau_{0},\vartheta\right)$ is the Jacobian for the transformation between the ray coordinates $\tau_{0}$,~$\vartheta$ and the Cartesian coordinates $y$,~$z$:
\begin{multline}
\mathcal{J}\left(\tau,\tau_{0},\vartheta\right)=
\left|\begin{array}{ccc}
\displaystyle\frac{\partial{y}\left(\tau_{0},\vartheta\right)}{\partial\tau_{0}},&
\displaystyle\frac{\partial{y}\left(\tau_{0},\vartheta\right)}{\partial\vartheta}\\
\displaystyle\frac{\partial{z}\left(\tau_{0},\vartheta\right)}{\partial\tau_{0}},&
\displaystyle\frac{\partial{z}\left(\tau_{0},\vartheta\right)}{\partial\vartheta}
\end{array}\right|{}\\{}=
R\textsl{g}\cos{\vartheta}\left(4\tau_{0}-3\tau\right)\!\bigl/2\bigr.+\textsl{g}^{2}\left(1+\cos{2\vartheta}\right)\tau_{0}^{2}\left(\tau-\tau_{0}\right){}\\{}-\textsl{g}^{2}\left(1+2\cos{2\vartheta}\right)\tau_{0}\tau\left(\tau-\tau_{0}\right)\!\bigl/2\bigr..
\end{multline}
This yields the expression for the change of density $a^{2}\left(y\left(\tau\right),z\left(\tau\right),\tau\right)$ along the characteristics~\eqref{eq:eq-yz-2} of the eikonal equation~\eqref{eq:eikonal-eq}:
\begin{equation}
a^{2}\left(y\left(\tau\right),z\left(\tau\right),\tau\right)=a^{2}\left(y_{0},z_{0},\tau_{0}\right)\frac{\mathcal{J}\left(\tau_{0},\tau_{0},\vartheta\right)}{\mathcal{J}\left(\tau,\tau_{0},\vartheta\right)},
\end{equation}
where $\mathcal{J}\left(\tau_{0},\tau_{0},\vartheta\right)\!=\!R\textsl{g}\tau_{0}\cos{\vartheta}\bigl/2\bigr.$. The zeros of $\mathcal{J}\left(\tau,\tau_{0},\vartheta\right)$ define the position of a spatio-temporal caustic, where two or more rays coalesce~\cite{MaslovFedoryuk_1981}. The position of the caustics at any given time $\tau$ is defined by the expressions~\eqref{eq:eq-yz-2}, with $\tau_{0}$ determined by the solutions of the following equation:
\begin{multline}\label{eq:CausticEq}
R\textsl{g}\cos{\vartheta}\left(4\tau_{0}-3\tau\right)\!\bigl/2\bigr.+\textsl{g}^{2}\left(1+\cos{2\vartheta}\right)\tau_{0}^{2}\left(\tau-\tau_{0}\right){}\\{}-\textsl{g}^{2}\left(1+2\cos{2\vartheta}\right)\tau_{0}\tau\left(\tau-\tau_{0}\right)\!\bigl/2\bigr.=0.
\end{multline}
This equation is cubic with respect to~$\tau_{0}$ and its solution looks cumbersome. However, we note that our theoretical approach works well enough under the condition $g\tau_{0}^{2}\!\gg\!{R}$. Therefore, one can neglect the first term in the left-hand side of Eq.~\eqref{eq:CausticEq} and find that $\tau_{0}$ and $\tau$ are connected approximately in a linear way
\begin{equation}\label{eq:caustic-t0-t}
\tau_{0}=\frac{\left(1+2\cos{2\vartheta}\right)}{2\left(1+\cos{2\vartheta}\right)}\tau.
\end{equation}
The value $\tau_{0}$ should be positive from its physical meaning. Hence, only the values $\vartheta$ from the interval ${-\pi\bigl/3\bigr.}\!\leq\!\vartheta\!\leq\!{\pi\bigl/3\bigr.}$ should be used for constructing the spatio-temporal caustics. We note also that the value $\delta{t}$ is much smaller than $t$ at a final stage of the discussed process. Hence, it can be considered that $\tau\approx{t}$ without any loss of accuracy. As a result, the form of the caustics in the plane $y$,~$z$ at an arbitrary time moment $t\gg{R^{2}}$ is determined by the expressions
\begin{subequations}\label{eq:characteristics}
\begin{gather}
y\left(\vartheta,t\right)=\frac{\sin{2\vartheta}\left(1+2\cos{2\vartheta}\right)}{8\left(1+\cos{2\vartheta}\right)^{2}}\textsl{g}t^{2},\label{eq:characteristics-y}{}\\{}
z\left(\vartheta,t\right)=\frac{\left(\cos{2\vartheta}+\cos{4\vartheta}\right)}{8\left(1+\cos{2\vartheta}\right)^{2}}\textsl{g}t^{2}.\label{eq:characteristics-z}
\end{gather}
\end{subequations}
Using this equations, it is possible to calculate the caustic curve corresponding to the maximum of the density $a^{2}\left(y\left(\tau\right),z\left(\tau\right),\tau\right)$ of atoms in the reflected atom laser wave packet by knowing only the minimal number of the experimental parameters.

The maximum density curve calculated using our analytical theory  (\ref{eq:characteristics}) is shown in Fig.~\ref{fig:caustic2}. It is clearly seen that, at longer evolution times, this curve defines the spatial position of the characteristic twin jets in the wake of the atom laser beam observed in the experiment (Fig.~2 (a)) and numerical simulations using the 2D GP mean-field model. 

\begin{figure}[t]
\centering
\includegraphics[width=8.6cm]{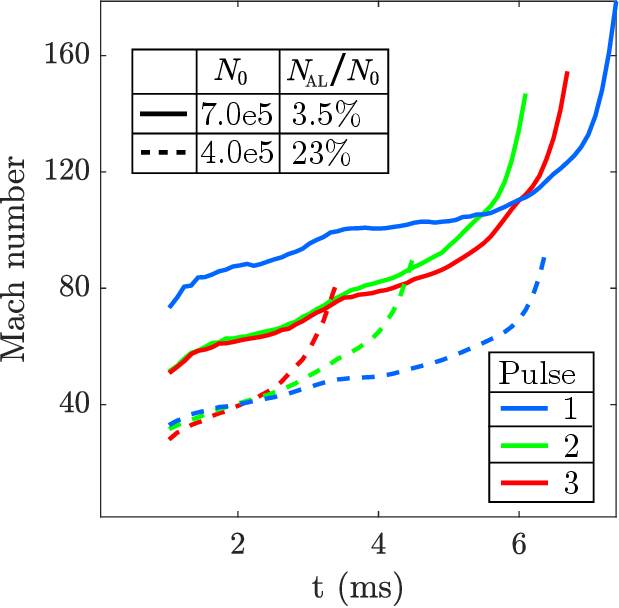}
\caption{(Color online) Development of hypersonic flow of the atom laser around the trapped BEC. $t=0$ indicates the time when the RF outcoupling pulse is applied. The dynamics of the Mach number at the BEC are illustrated for two different experimental configurations: $N_{0}=7.0(3) \times 10^{5}$ and $N_{\textrm{AL}}/N_{0}=3.5(3)\%$ (solid line), and $N_{0}=4.0(4) \times 10^{5}$  and $N_{\textrm{AL}}/N_{0}=23(2)\%$ (dashed line). The local Mach numbers are only shown in times between the formation of the atom laser and while  regions it is still in contact with the BEC. 
}
\label{fig:mach_dyn}
\end{figure}

\section{Experimental results}
The experimental results for a BEC with an initial (before atom laser) atom number of $N_0 = 7.0(3) \times 10^5$ and an outcoupling fraction of $N_{\textrm{AL}}/N_{\textrm{0}} = 3.5(3)\%$ are shown in Fig.~\ref{fig:RFALShocks}(a) and agree well with the corresponding numerical simulation (see Fig.~\ref{fig:RFALShocks}(d)). In both cases the jets resulting from the BEC exiting the atom laser are clearly visible. The simulations also predict the characteristic cross-hatching pattern caused by the BCR which results from the transit of the BEC through the atom laser beam. 

When the number of atoms in the initial BEC is significantly reduced (see Fig.~\ref{fig:RFALShocks}(b)), the fringe pattern remains essentially the same, although the atom laser is significantly smaller. However, if the outcoupling fraction is greatly increased to $N_{\textrm{AL}}/N_{0}=67(2)\%$, the fringes wash out (see Fig.~\ref{fig:RFALShocks}(c)).

To compare our experimental results to simulation and define the Mach numbers present in our system we note that the near-field spatial profile can be approximated by taking the velocity of atoms in the far-field from the measured spatial profile and dividing by the time-of-flight expansion time. From this estimated near-field density profile $n$, it is possible to find the speed of sound, $c$, for the appropriate regions in the atom laser at the time of interaction with the BEC. The speed of sound in a weakly interacting Bose gas is given by $c=\sqrt[]{U_{aa}n/m}$, where $U_{aa}=4\pi\hbar^2a_{0}/m$ is the mean-field interaction strength for atoms in the atom laser, for which we use the particle $s$-wave scattering length $a_\textrm{0}=5.56~\textrm{nm}$ \cite{PhysRevA.64.042710} as all atoms are in the $m_J=0$ state. We estimate the relative velocity of the BEC with respect to the atom laser as ${\sim}13 \textrm{-} 41$~mm/s, which greatly exceeds the local speed of sound of ${\sim} 0.1 \textrm{-} 0.5$~mm/s.

Fig.~\ref{fig:mach_dyn} shows the approximate evolution of this Mach number in our experiments of a freely falling atom laser beam. 
Indeed, most of the interaction between the atom laser and the trapped BEC happens in the high-hypersonic regime (Mach number greater than 10), while a significant portion in the trailing tail of the atom laser for some experimental configurations are higher still (Mach number $\gtrsim 80$). 

Increasing the number of atoms in the BEC increases the maximum flow velocity, while increasing the number of atoms in the atom laser increases the local speed of sound, and vice versa, which produces opposite effects on the Mach number. 
In practice, the Mach regime is experimentally controllable by adjusting the population of the initial condensate and the RF-outcoupled fraction in opposite directions (see Fig.~\ref{fig:mach_dyn}).

The characteristic disk-like profile of the RF atom laser is well explained by considering only the mean-field interaction between the out-coupled atoms and the trapped BEC (see Appendix~\ref{apx_det_atom_laser_size}).
Here we present a quantitative analysis of the anomalous interference pattern observed in the far-field profile of the atom laser in light of the numerical simulation and analytic theory and show that the peculiar fringe spacings are well explained by BCR theory.

\begin{figure}[t]
\centering
\includegraphics[width=8.6cm]{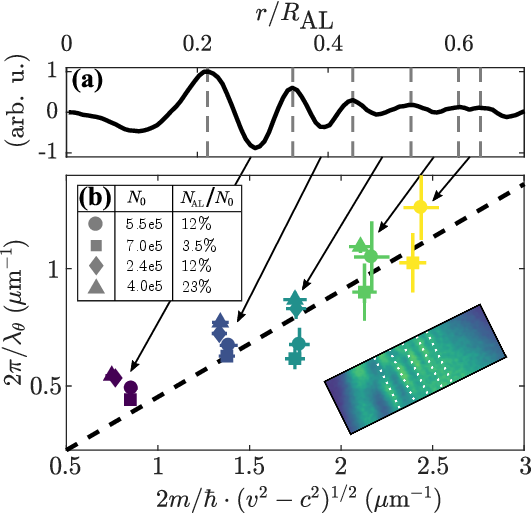}
\caption{(Color online) (a) Density modulation in the atom laser. The solid line is the 1D density profile along the right jet of the atom laser shown in Fig.~\ref{fig:RFALShocks}(a) with a band-pass filter applied to remove the slow change in density due to the atom laser bulk shape and reduce shot noise. Dashed vertical lines mark the series of peaks along the profile.
(b) Fringe spacing in BCR. 
Markers are the experimental data: colours index adjacent pair of fringes from the series of emanating wave fronts; shapes distinguish experimental configurations ($N_0$,$N_{\textrm{AL}}/N_{0}$). 
The dashed line is the approximate prediction from BCR theory, for which the spacing between wavefronts are taken along the jet angle, consistent with the analysis of the experimental data. Inset shows the region (Fig.~\ref{fig:RFALShocks}(a)) integrated (along the short axis) to produce (a), dashed lines indicate fringe positions. Fringe alignment with the short axis of the region is apparent.
}
\label{fig:data_BCR_theory}
\end{figure}

Fig.~\ref{fig:data_BCR_theory}(a) shows a line profile of the density in the far-field of the atom laser taken along the characteristic jet.
According to BCR theory, the periodicity of the density peaks in the early dynamics are directly related to the flow velocity and the speed of sound in the atom laser \cite{CarusottoHuCollinsEtAl2006}. The non-equal spacing between between two adjacent   wavefronts demonstrates the impurity's (effective) acceleration. 
More specifically, BCR is characterised by a fan-shaped series of conical wavefronts with a density oscillation at a wavenumber given by $k_{\textrm{BCR}} = 2\pi/\lambda = 2m/\hbar \cdot (v^2 - c^2)^{1/2}$, measured directly upstream of the flow.

According to the GP simulations of our experiment, a fan-shaped series of parabolic wavefronts in the upstream of the condensate detach from the bulk shortly after they are generated (see Fig.~\ref{fig:TheoryTimeSeries}(b,c)). 
Therefore, a simple uniform scaling will transform the density oscillations measured along the jets in far-field into the relevant near-field counterpart.
Thus we estimate the physically relevant parameters at the early dynamics, such as the local flow velocity ($v$), speed of sound ($c$), and period of density modulation ($\lambda_{\theta}$, where $\theta$ specifies angle relative to upstream), using the far-field profile measured at the detector (see Fig.~\ref{fig:RFALShocks}).

Fig.~\ref{fig:data_BCR_theory}(b) shows the calculated density modulation at the early dynamics and the prediction from the BCR theory approximated at the jet angle (${\sim} 60^{\circ}$ with respect to the upstream direction).
The quantitative agreement between the experimental result and the theory offers a strong evidence for BCR as the mechanism for the originator of the bright interference pattern.

\begin{figure}[t]
\centering
\includegraphics[width=8.6cm]{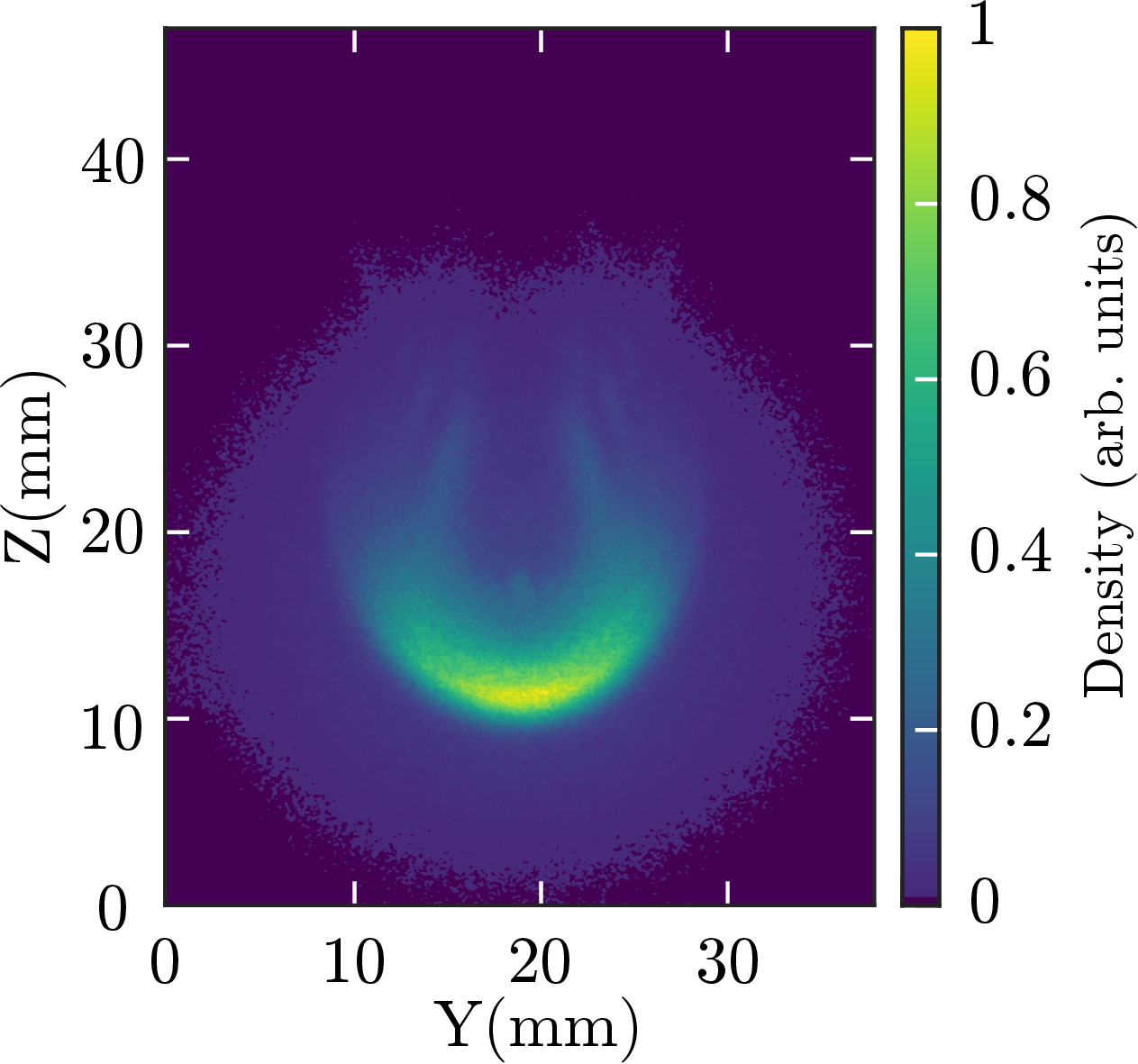}
\caption{(Color online) For weaker trap frequencies ($\{\omega_x,\omega_y,\omega_z\}=2\pi\ ( \{57,227,234\}) ~\textrm{Hz}$) and high atom number ($N_0=1.0(3)\times10^{\textrm{6}}$, $N_{\textrm{AL}}/N_{0}=8(3)\%$) the outcoupling mostly occurs in the downward direction due to the decreased mean field potential.
}
\label{fig:low_trap_freq}
\end{figure}

To provide an intermediate case between the complex spatial profiles seen in our main experiments and the far simpler bean-shaped profiles observed elsewhere \cite{PhysRevLett.78.582,Anderson1686}, we repeat the experiment with lower trapping frequencies. A profile of an atom laser formed from a $\{\omega_x,\omega_y,\omega_z\}\approx2\pi\{57,227,234\}~\textrm{Hz}$ trap is shown in Fig.~\ref{fig:low_trap_freq}. Here the lower confinement(and in turn mean field) causes atoms to be directed  downwards, rather then than expand in all directions \cite{Dall2007}. This leads to fainter BCR signatures eg. the fringing and jets are much less visible in the profile. 

\section{Conclusions\label{sec:conclusions}}

In conclusion, we have demonstrated a simple but powerful system for studying BCR using an atom laser.
The BCR results from the high-hypersonic motion of the BEC through the atom laser, analogous to the motion of a barrier through a quantum fluid.

Our results are in excellent agreement with mean-field numerical simulations, as well as with a comprehensive analytic theory. This work has strong implications from the viewpoint of using atom lasers for interferometry since the BCR acts to degrade the beam quality of the atom laser and thus potentially limit its ultimate sensitivity in interference measurements.

These experiments could be further extended by using a Feshbach resonance to tune the interactions in an atom laser beam and investigate the effect of BEC barrier height/size on the resulting dynamics. In such case one should be able to study non-linear dynamics such as shock waves, soliton trains and vortices \cite{Carusotto2013}.

Furthermore, due to the accelerating high-hypersonic regime demonstrated in our experiments this system may be used to study condensed matter analogs of relativistic electrodynamic effects, such as the dynamic Casimir force, quantum friction \cite{PhysRevLett.118.045301} and the Unruh effect \cite{Unruh:1976ey,Unruh:1981hs,Retzker:2008bp} along with analog gravity effects, such as Hawking radiation \cite{PhysRevA.70.063602,PhysRevLett.94.061302}.

\begin{acknowledgments}
We thank D. Cocks for careful reading of the manuscript.
This work was supported through Australian Research
Council (ARC) Discovery Project Grant No. DP160102337. X. Y. is supported by NSFC 11504328.
D.K.S. is supported by an Australian Government Research Training Program (RTP) Scholarship.
S. S. H. is supported by ARC Discovery Early Career Researcher Award No. DE150100315.
L. A. S. is supported by the Russian Science Foundation (Grant 17-12-01534). X. W. G. is supported by  the key NSFC grant No.\ 11534014 and the National Key R\&D Program of China  No. 2017YFA0304500.
\end{acknowledgments}

\appendix

\section{Atom Laser Size}\label{apx_det_atom_laser_size}
\begin{figure}[t]
\centering
\includegraphics[width=8.6cm]{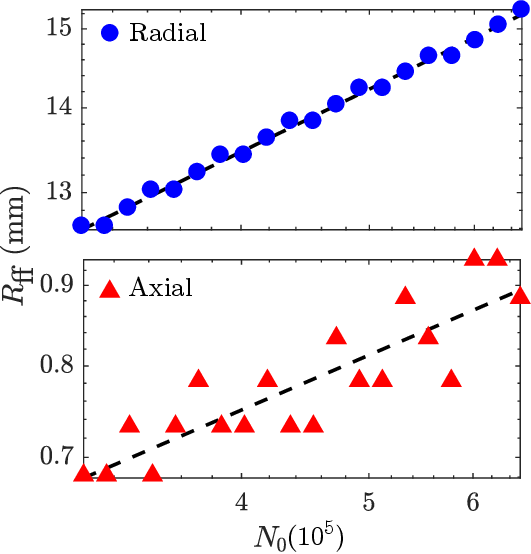}
\caption{(Color online) Characteristic size scaling of the pulsed atom lasers in the far-field. The bulk size of the atom lasers is characterised by the profile's half-width at half-maximum ($R_\textrm{ff}$). Circles and triangles represent the sizes measured along the radial (perpendicular) and axial dimensions of the cigar trap, respectively. The dashed lines are the scaling law fits to data given by $R_{\textrm{ff}} \propto N_0^{1/5}$, which is derived from assuming the kinetic energy of outcoupling atoms in the atom laser being proportional to the condensate chemical potential. The data shown is an average of approximately 4000 experimental runs of 20 consecutive pulsed atom lasers at $N_{\textrm{AL}}/N_{0}=3.5(3)\%$, with initial BEC  containing $N_0=7.5(3)\times10^{\textrm{5}}$. The reader should note the logarithmically spaced axes.
}
\label{fig:al_size_scaling}
\end{figure}

The bulk size and shape of the pulsed atom lasers in the far-field is due to the initial mean-field repulsion from the trapped BEC and the subsequent ballistic expansion, with minimal effects from interactions within the atom laser, where the density is low. As expected from the simple conversion of the mean-field chemical potential ($\mu \propto N_0^{2/5}$) into velocity ($v\propto \mu ^{1/2}$), the sizes of the atom lasers ($R_{\textrm{ff}}\propto v$) in the far-field scale according to $R_{\textrm{ff}} \approx \sqrt{\mu} \propto N_0^{1/5}$ (see Fig.~\ref{fig:al_size_scaling}). 

\section{Detector Temporal Resolution}\label{apx_det_res}
Although the ultimate timing resolution for an MCP-DLD system is of order nanoseconds, this resolution is only realized in the case of fast moving particles. In the case of slow moving particles, the time delay between a particle striking the top of the MCP angled pore and a particle striking deep in the pore becomes the limiting temporal effect.

In the worst case this delay will be $t_{\textrm{max}}=2 r_{\textrm{pore}}/\sin{(\theta)}v$ where $\theta$ is the pore angle to vertical, $r_{\textrm{pore}}$ is the pore radius and $v$ is the incident speed to the detector. 
For our system with $r_{pore}=5$~\textmu{}m, $\theta=12^{\circ}$ and $v{\sim} 4$~m/s  this gives $t_{max} =12$~\textmu{}s.

Assuming a uniform flux at the surface and the propagation time along the pore for electrons produced from a He* impact is negligible, the probability distribution function of of the delay $t_{\textrm{delay}}$ relative to the top of the pore can be found from simple geometric arguments to be:
\begin{equation}
f(t_{\textrm{delay}})=\frac{4 \sqrt{t_{\textrm{max}}^2-t_{\textrm{delay}}^2}}{ \pi  t_{\textrm{max}}^{2}},
\label{eq:det_dist}
\end{equation}
which has a standard deviation of
\begin{equation}
\sigma_{delay}=t_{\textrm{max}} \frac{\sqrt{9 \pi ^2-64}}{6 \pi }\approx0.26 t_{\textrm{max}}.
\label{eq:det_std}
\end{equation}
For our system this yields a spread of $\sigma_{delay}\approx 3.1$~\textmu{}s, which is an estimate of our temporal resolution. 


\section{2D Gross Pitaevskii Simulations - Vortex Shedding}\label{apx_gp_sim}
\begin{figure}[h]
\includegraphics[width=8.6cm]{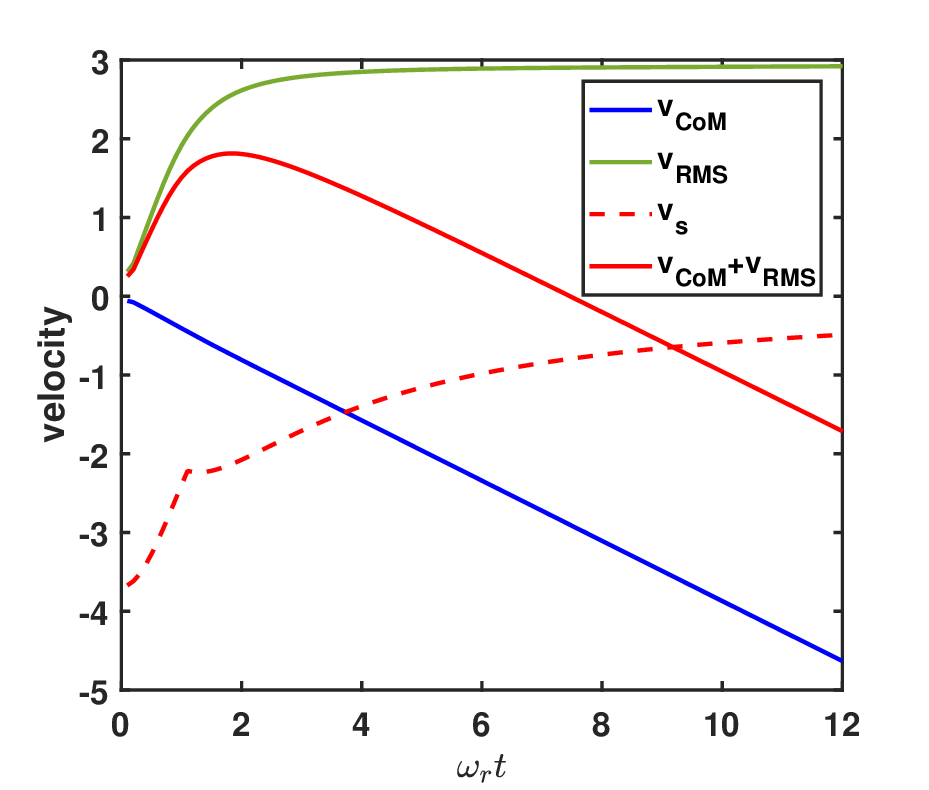}
\caption{(Color online) Comparison of relevant velocities in the experiment as a function of time. $V_\textrm{COM}$ is the velocity of the centre-of-mass of the atom laser. $V_\textrm{RMS}$ is the root-mean-squared width of the atom laser field. $V_\textrm{s}$ is local speed of sound.}
\label{fig:Velocity}
\end{figure}
The dynamics of the atom laser after the outcoupling RF pulse is switched off can be divided into three stages. At the initial stage, the center-of-mass (CoM) velocity of the atom laser remains small, the atom laser and the BEC strongly overlap, and the displacement between their CoM positions is small. The cloud of outcoupled atoms rapidly expands due to the strong repulsive interaction with the condensate. At the end of this stage, it forms a concentric shape as shown in Fig.~1(a).

The velocity that determines the BCR is the overall velocity of the atom laser atoms upstream from the condensate. It can be approximated by the sum of the CoM velocity which is directed downstream and the expansion velocity which is directed upstream for the upper part of the atom laser. The latter can be estimated by the changing rate of the root-mean-squared width of the atom laser field. Fig.~\ref{fig:Velocity} shows a comparison of the velocity difference and the sound velocity as a function of time. At the end of the first stage of the evolution, these two velocities add up to zero. The duration of this stage (${\sim} 1-2$~ms) depends on the initial atom number in the condensate, which determines the strength of the repulsive barrier, and hence the expansion velocity. After that, the upper part of the atom laser starts to flow downwards and the dynamics enters the second stage, during which the interaction between the atom laser and the repulsive barrier results in  very rich dynamics.
\begin{figure*}
\centering
\includegraphics[width=500px]{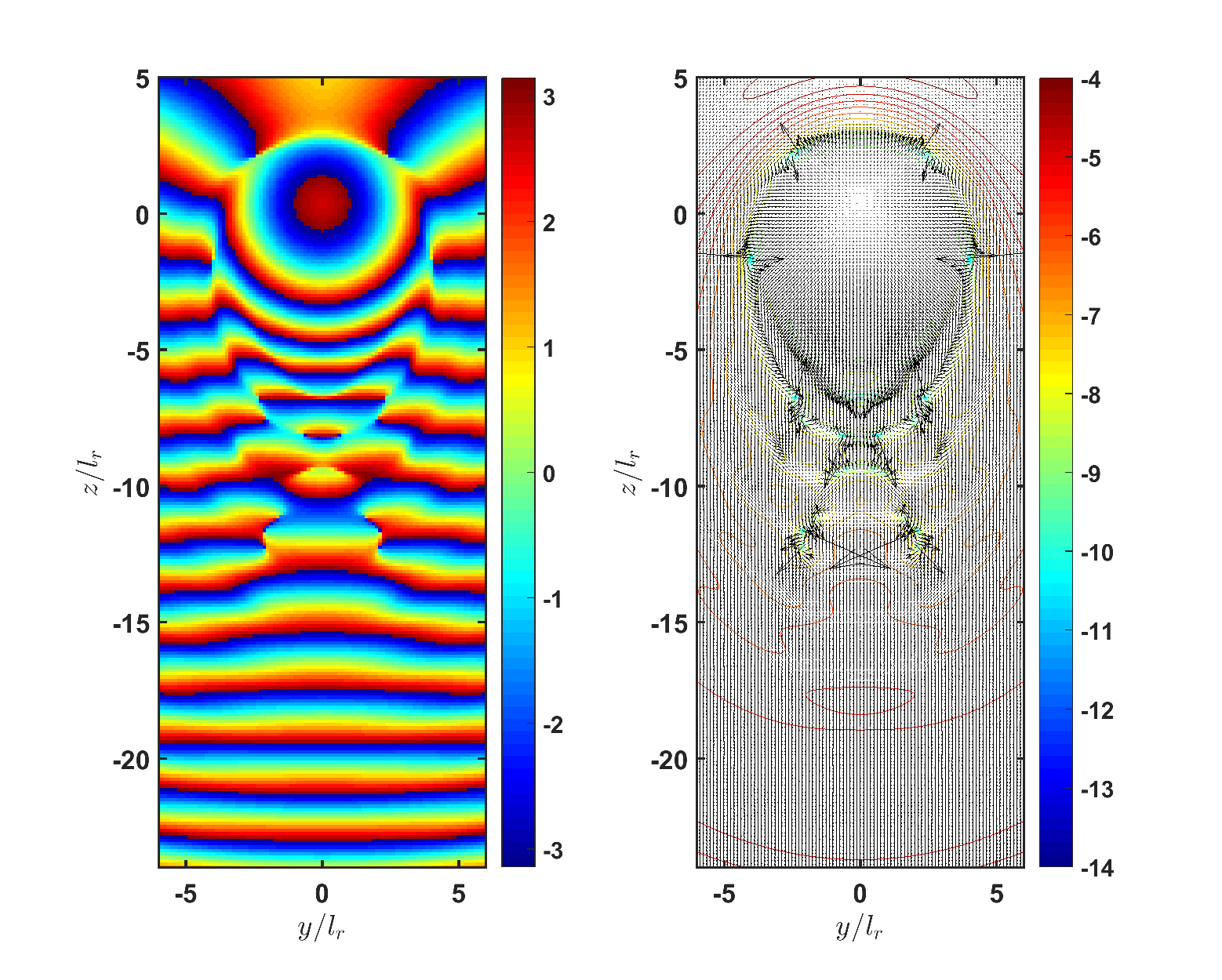}
\caption{(Color online) The phase of the wave function and quiver plot of velocity showing vortices pairs in the vicinity
of the repulsive barrier at $\omega_\textrm{r} t=9$. The colored lines are contours of the density.}
\label{fig:Vortices-pairs}
\end{figure*}

One of the most interesting processes in the beginning of the second stage is the emergence of vortices in the vicinity of the barrier. In the free falling reference frame, the potential barrier accelerates upward in the field of the atom laser. Once local relative velocity exceeds the sound velocity in the atom laser cloud (at $\omega_\textrm{r} t {\simeq} 9$ in Fig.~\ref{fig:Velocity}), the perturbation caused by the barrier cannot expand across the whole field in time, and the density near the barrier fluctuates heavily over a small length scale. To reduce this extra generated kinetic energy, vortex pairs of opposite sign are nucleated and drift downstream. This can be seen from the phase dislocations in the phase plot of the wave function as well as in the  quiver plot of the atom laser velocity field in Fig.~\ref{fig:Vortices-pairs} with the logarithm contour of the density. The density at the positions of vortices is zero (the blue contours correspond to $10^{-12}$). While the vortex pairs are also generated at later times, the ones generated initially do not disappear but flow downstream to form the so-called ``vortex street''. In contrast to previous results \cite{barenghi_primer_2016,Mironov2010,PhysRevLett.113.103901,PhysRevLett.115.089401,Kamchatnov2015} here the vortex pairs are not equally spaced due to the acceleration under gravity. 

These vorticies are unlikely to be observed experimentally for two reasons: First the spatial extent of the density disruption around the vortex is localized to a vanishingly small region in the far field, well below the detector resolution. Second as the experimental profiles are the summation of many thousand experimental realizations small changes in the experimental conditions may lead to a blurring that destroys such a feature. 

The process described above lasts until the barrier passes through the
upper part of the atom laser field, as seen in Fig.~1(d). After this, the atom laser beam can been treated as free falling since it is now far away from the
barrier and is dilute enough to ignore the interaction between 
the atoms. The density distribution at long expansion times coincides with the
instantaneous momentum distribution, as shown in Fig.~2(d).

\newpage
\bibliographystyle{apsrev4-1}
\bibliography{bcral}

\end{document}